\documentclass[twocolumn]{aastex631}
\usepackage{color}
\usepackage{amsmath}
\usepackage{CJK}

\newcommand{\Ha}{\makebox{H$\alpha$}}

\newcommand{\hbeta}{H{$\beta$}}
\newcommand{\halpha}{H{$\alpha$}}

\def \OIII {[O\,{\sc iii}]}

\newcommand{\SII}{[S{\sevenrm\,II}]}

   \font\sevenrm=cmr7 scaled 1000

\newcommand{\FeII}{Fe{\sevenrm\, II}}

\usepackage{soul}
\usepackage{changepage}
\usepackage{multirow}

\begin{document}
\begin{CJK*}{UTF8}{gbsn}

\title{Too Quiet for Comfort: Local Little Red Dots Lack Variability over Decades}

\author[0000-0002-8501-3518]{Colin J. Burke}
\affiliation{Department of Astronomy, Yale University, New Haven, CT 06511, USA}

\author[0000-0002-8501-3518]{Zachary Stone}
\affiliation{Department of Astronomy, University of Illinois at Urbana-Champaign, Urbana, IL 61801, USA}

\author[0000-0003-1659-7035]{Yue Shen}
\affiliation{Department of Astronomy, University of Illinois at Urbana-Champaign, Urbana, IL 61801, USA}
\affiliation{National Center for Supercomputing Applications, University of Illinois at Urbana-Champaign, Urbana, IL 61801, USA}

\author[0000-0002-2624-3399]{Yan-Fei Jiang (姜燕飞)}
\affiliation{Center for Computational Astrophysics, Flatiron Institute, New York, NY 10010, USA}


\begin{abstract}
Several local ($z\lesssim 0.2$) metal-poor dwarf AGNs have remarkably similar properties to those of high-redshift Little Red Dots (LRDs), and are recently proposed to be the local analogs of LRDs. We use long-term photometric and spectroscopic observations of three local LRDs spanning $\sim 20$ years to measure variability in their rest-frame optical continuum and broad \halpha\ emission lines. Using ZTF light curves over a rest-frame $\sim 5$~yr baseline, the $r$-band intrinsic rms variability is $(9\times 10^{-5})_{\rm -9E-5}^{+0.014}$ mag (J1022), $0.025\pm0.004$ mag (J1025) and $0.020\pm0.005$ mag (J1047), indicating low intrinsic variability ($<3-4\%$ at 3$\sigma$). These rms variability amplitudes are much lower than those for dwarf AGNs and more massive quasars. There is little structure in the optical variability structure functions for the three local LRDs, in contrast to normal AGN variability. Using available multi-epoch spectra, we constrain the broad \halpha\ line flux variability to be less than a few percent, without significant profile changes, over a rest-frame baseline of $\sim 15$~yrs in J1025 and J1047, respectively. The three LRDs stand out in the Balmer line properties compared with normal broad-line AGNs, with exceptionally large \halpha\ equivalent widths and \halpha/\hbeta\ ratios far exceeding the Case B recombination value. In the context of recent theoretical models of LRDs as dense gas-enshrouded massive black holes with super-Eddington accretion, our results suggest that the photosphere emission is long-term stable and the broad Balmer lines are primarily collisonally excited. This scenario is consistent with the lack of variability, large \halpha/\hbeta\ ratios and little dust extinction, as well as the expected high gas density. Virial black hole mass estimates using broad \halpha\ assuming photoionization are therefore highly questionable for LRDs. 

\end{abstract}





\section{Introduction} \label{sec:intro}

In search for metal-poor dwarf emission-line galaxies in the low-redshift Universe, \citet{Izotov_Thuan_2008} identified four metal-poor dwarf galaxies with broad emission lines that are indicative of AGN activity \citep[e.g.,][]{Izotov_Thuan_2008,Simmonds_etal_2016,Burke_etal_2021}. Long-term repeated spectroscopy shows that the broad emission lines in these systems are long-lived over a decade \citep{Simmonds_etal_2016,Burke_etal_2021}, ruling out a stellar transient origin for the broad lines. If these metal-poor broad-line emitters are AGNs, their bolometric luminosities estimated from rest-frame optical continuum are approximately {$10^{44}-10^{45}\,{\rm erg\,s^{-1}}$}, placing them in the low-luminosity AGN regime. Their black hole masses estimated from the broad Balmer lines are typically $\sim 10^{6-7}\,M_\odot$. These metal-poor dwarf AGN candidates have compact morphology in rest-frame optical, and weak X-ray emission with no robust detection \citep[e.g.,][]{Simmonds_etal_2016}. Long-term monitoring also shows that these metal-poor dwarf AGN candidates have weak to none optical variability \citep{Burke_etal_2021}. 

On the other hand, JWST observations have revealed a remarkable population of high-redshift ($3\lesssim z\lesssim 9$) systems with broad emission lines, compact morphology and red colors in rest-frame optical, dubbed Little Red Dots (LRDs) \cite[e.g.,][]{Matthee+24,Greene+2024,Kocevski_LRD_selection}. Exceptionally strong Balmer breaks are often observed for high-redshift LRDs \citep[e.g.,][]{Wang_etal_2024,Ji+2025,Naidu2025}. They are generally X-ray weak \citep{Yue_etal_2024,Maiolino2025}, and lack detectable optical continuum variability \citep{Kokubo2024,ZhangEtAl2025,Stone_etal_2025}, and have weak hot dust emission characteristic of an AGN torus \citep[e.g.,][]{Setton_etal_2025}. While the nature of LRDs remains unclear, recent theoretical models suggest that most properties of LRDs can be attributed to a dense gas-enshrouded supermassive black hole, potentially accreting at high Eddington ratios \citep[e.g.,][]{Naidu2025,LiuEtAl2025}. 

\citet{Lin_etal_2025_egg} recently made an important connection between low-$z$ metal-poor dwarf AGNs and high-$z$ LRDs \citep[also see][]{Ji2025}, noting the remarkable similarities between the two populations: (1) both populations have similar spectral energy distributions, featuring a red rest-frame optical continuum, a ``V''-shaped upturn into rest-frame UV, and weak dust emission at rest-frame few microns; (2) Balmer absorption features are often observed in both populations, indicative of high gas densities; (3) both populations show compact morphology in rest-frame optical, indicating a nuclear origin for the optical continuum; (4) both populations show weak X-ray emission and optical variability. \citet{Lin_etal_2025_egg} further suggested that these properties can be explained in a unified scheme of LRDs, where the central accreting SMBH is embedded in a dense atmosphere formed in super-Eddington accretion, based on the theoretical model in \citet{LiuEtAl2025} -- also see the ``quasi-star'' model proposed by \citet{Begelman_etal_2008} that bears some conceptual similarities with the super-Eddington accretion atmosphere model \citep[also see][]{Begelman_Dexter_2025}. The atmosphere forms a photosphere at $\sim 10^{4}-10^{5}r_g$ with an effective temperature of $\sim 5000$~K to emit the observed rest-frame optical SED. The broad-line emission comes from the polar region of the photosphere, where leaked ionizing flux from the accreting SMBH powers the observed broad emission lines. Alternatively, the broad emission lines could be produced by hydrogen scattering and electron scattering in the dense gas envelope, as suggested by the ``black hole star'' model (hereafter BH*) proposed by \citet{Naidu2025}. Either way, these gas-enshrouded models represent by far the most promising models to explain the peculiar properties of LRDs. 

Although LRDs become increasingly rare towards low redshift \citep[e.g.,][]{Ma_etal_2025}, the low redshifts and brightness of local LRD analogs make them excellent targets for detailed studies to test the above theoretical models for LRDs. Variability constraints, for example, can offer valuable insights on the nature of the emitting regions \citep[e.g.,][]{Inayoshi2024xray,Zhou_etal_2025,SecundaEtAl2025}. In this work, we investigate the long-term variability properties of the three local metal-poor dwarf AGNs (denoted as local LRD analogs) reported in \citet{Lin_etal_2025_egg}, J1022, J1025 and J1047. The three local LRDs have estimated virial BH masses of $10^6-10^7\,M_\odot$ \citep{Lin_etal_2025_egg}. With past observations of these objects more than 15 years ago, we sample rest-frame timescales of at least 10 years of the local LRD analogs. This is a distinct advantage compared with repeated JWST observations of high-$z$ LRDs, which only sample rest-frame timescales of less than a year \citep[e.g.,][]{Kokubo2024,ZhangEtAl2025,Stone_etal_2025}. Much longer rest-frame baselines can only be achieved for high-$z$ LRDs if they are also gravitationally lensed \citep[e.g.,][Zhang et al., in prep]{Furtak2025}, where lensing time delays lead to different snapshot epochs of individual images. However, systematics in lens modeling may introduce artificial flux variations across individual lensed images, and complicate the intrinsic variability measurements. 

We collect public photometric light curves and multi-epoch spectra for the three local LRDs in \citet{Lin_etal_2025_egg}. While Lin~et~al. briefly discussed variability properties for these systems, here we perform a more quantitative and complete analysis of their photometric and spectral variability, and comparisons with normal AGNs. The structure of this paper is as follows. In Section~\ref{sec:data} we describe the compiled public data, including our own observations of these objects in the past. In Section~\ref{sec:results} we present our main results and comparisons with a control AGN sample. We discuss the implications of our observations in Section~\ref{sec:disc} and conclude in Section~\ref{sec:con}.

\section{Data}\label{sec:data}

\begin{figure}
    \centering
    \includegraphics[width=\linewidth]{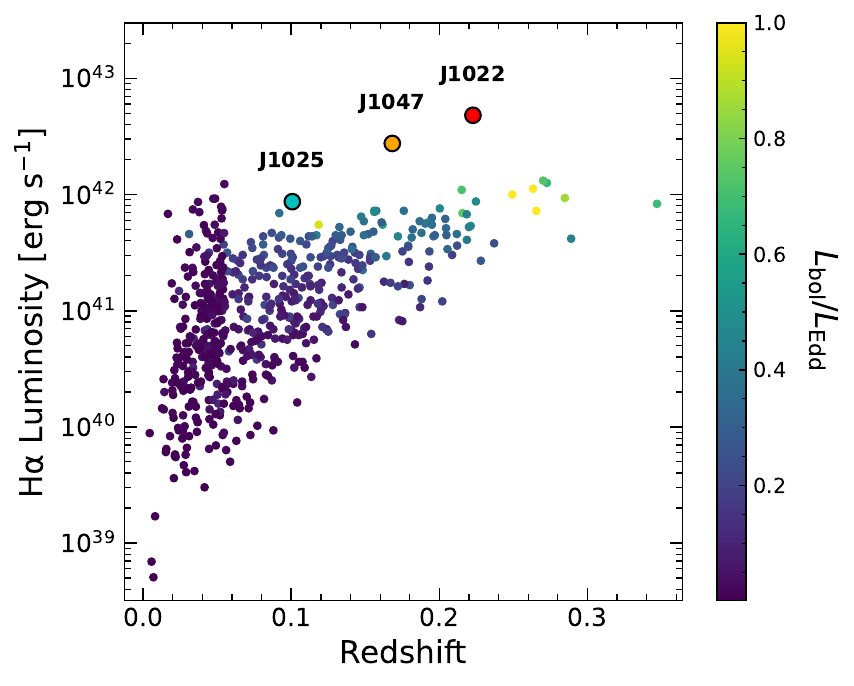}
    \caption{The \halpha\, luminosity against redshift for the dwarf AGN sample and the three local LRDs (labeled). Each dwarf AGN is colored by its Eddington ratio. The dwarf AGNs have \halpha\ FWHM $\gtrsim 1000\ {\rm km\,s^{-1}}$.  }
    \label{fig:Lha_v_z}
\end{figure}


We adopt the sample of local spectroscopically-selected broad-line AGNs from \citet{Wang2023} as our comparison sample for continuum variability. The \citet{Wang2023} sample comprises local AGNs from \citet{Reines2015} and additional dwarf AGNs from \citet{Greene2007,Chilingarian2018,Liu2018}. These dwarf AGNs were selected based on the presence of AGN-like narrow emission line ratios and broad Balmer emission lines in SDSS. {To ensure that our comparison sample are truly broad-line dwarf AGNs, we remove sources from \citet{Chilingarian2018}, as their black hole masses are low ($\log_{10}(M_{\rm BH}/M_{\odot}) \lesssim 5.3$) and contain narrower \halpha\,  (FWHM $\lesssim$ 1000 km s$^{-1}$).} Our comparison sample of 573 objects spans black hole masses of $\sim 10^5 - 10^8 M_{\odot}$ and $z<0.35$. The virial black hole masses of our comparison sample straddle the range of our local LRDs (if the LRDs virial black hole masses are taken at face value), allowing us to compare their variability and broad-line properties.

\begin{figure}[!t]
\centering
  \includegraphics[width=0.48\textwidth]{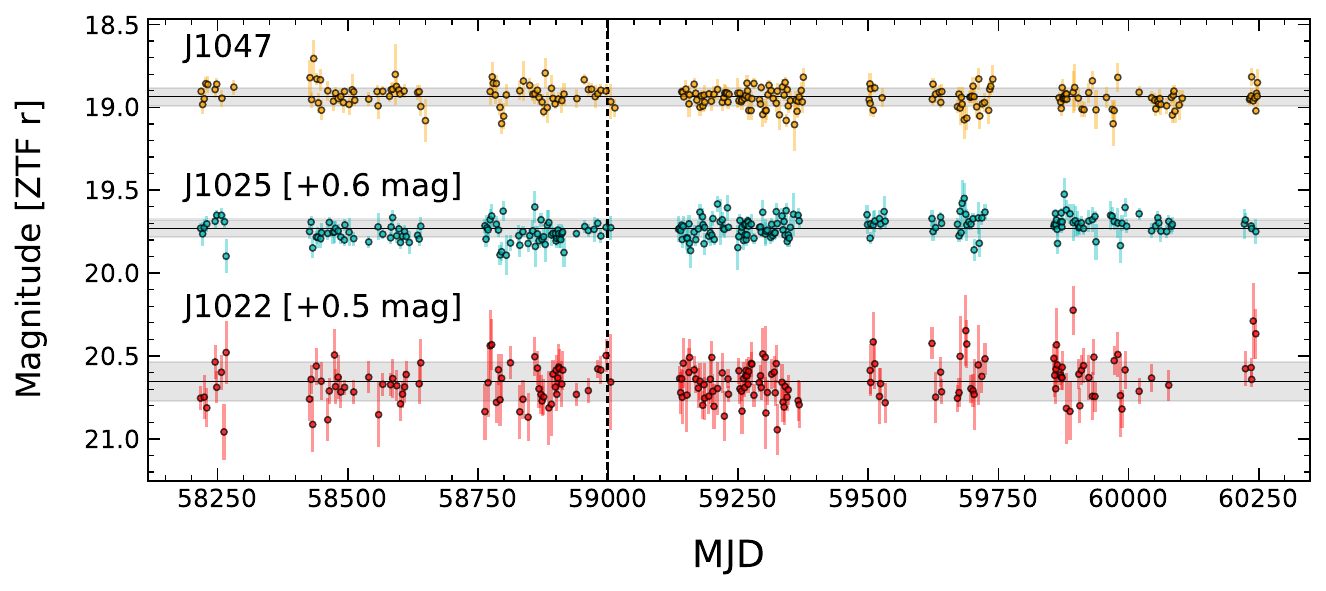}
  \caption{ Optical ZTF \emph{r}-band light curves for the three local LRDs in our sample. J1022 and J1025 have been shifted by an amount labeled above their light curves. Each light curve displays the mean (black line) and the mean measurement uncertainty (shaded region around the mean). The vertical dashed line indicates when Gemini spectra were obtained for J1025 and J1047. }
  \label{fig:lc}
\end{figure}

\begin{figure}[!t]
\centering
  \includegraphics[width=0.48\textwidth]{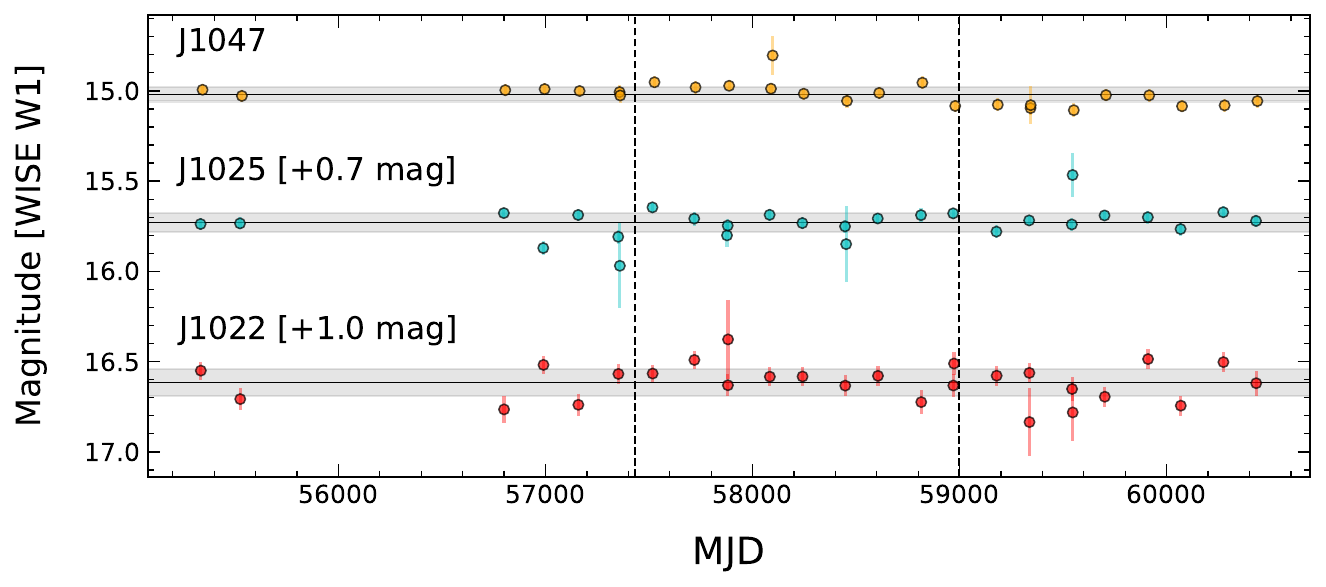}
  \caption{Infrared WISE W1 light curves for the three local LRDs in our sample. J1022 and J1025 have been shifted by an amount labeled above their light curves. Each light curve displays the mean (black line) and the mean measurement uncertainty (shaded region around the mean). The vertical dashed lines indicate when Keck and Gemini spectra were obtained for J1025 and J1047.}
  \label{fig:lc2}
\end{figure}

We utilize public photometric light curves (PSF magnitude) from the Zwicky Transient Facility \citep[ZTF;][]{ZTF}, provided by the Zubercal\footnote{http://nesssi.cacr.caltech.edu/ZTF/Web/Zuber.html} photometry service. Zubercal uses improved photometric calibrations to provide re-calibrated ZTF light curves up to its most recent release in February 2025. We query Zubercal for the three local LRDs and our dwarf AGN comparison sample with a matching radius of 3\arcsec. We bin the photometric light curves on a nightly basis, where we take the inverse-variance-weighted mean of all intra-night measurements (Fig.~\ref{fig:lc}). Similarly, we compile infrared WISE W1 and W2 band light curves for the three local LRDs and comparison sample using a search radius of 3\arcsec, binning nightly (Fig.~\ref{fig:lc2}).

\begin{figure*}
    \centering
    \includegraphics[width=\linewidth]{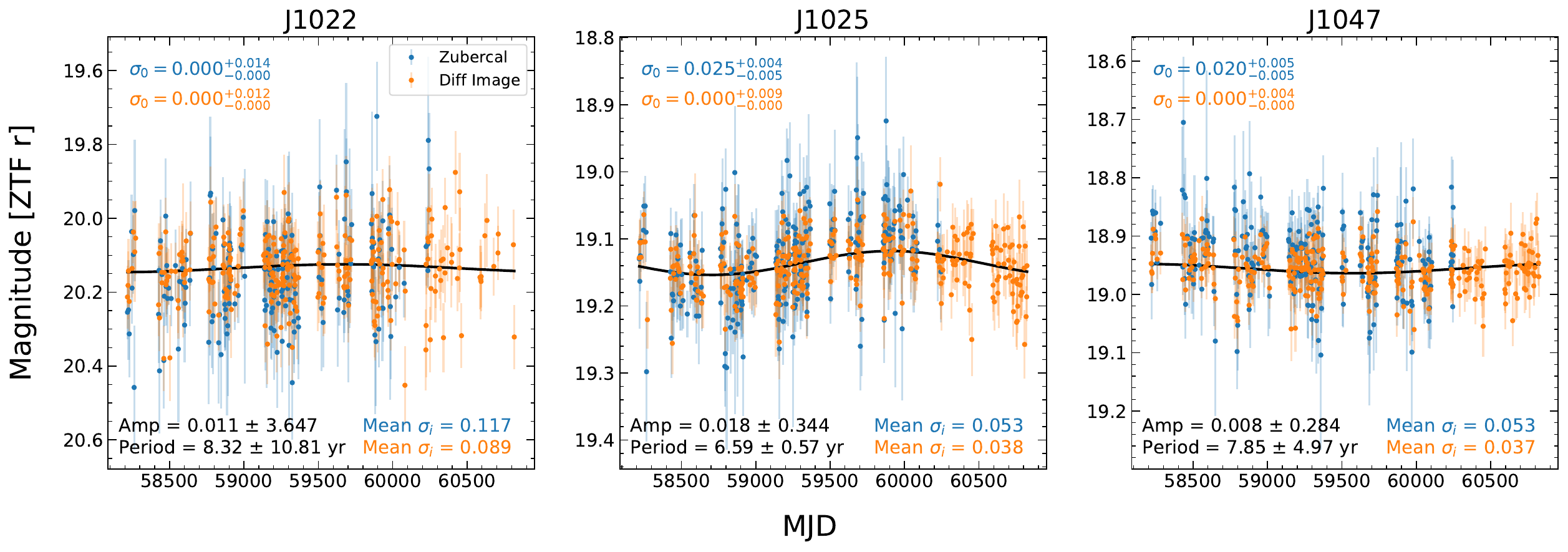}
    \caption{A compairson of the Zubercal and difference image \emph{r}-band light curves for the local LRDs. For each LRD, the intrinsic variability estimate $\sigma_0$ and mean measurement uncertainty $\overline{\sigma_i}$ are shown for each method. Sinusoidal fits to the difference image light curves are shown in black. The best-fit sinusoidal amplitude and period are shown in the bottom left.}
    \label{fig:diff_image_lc}
\end{figure*}

While the local LRDs are unresolved, the comparison dwarf AGN sample are mostly resolved, with much of the sample having extents $>$2\farcs5 (see \citet{Wang2023} Fig.~1). This induces additional scatter in the ZTF light curves due to improper PSF matching, and can artificially increase the observed variability. As a sanity check on our continuum variability analysis, we follow a similar process as \citet{Wang2023}, and construct ZTF light curves with aperture photometry on all difference images provided by the ZTF archive for our LRDs and the dwarf AGNs. The ZTF difference images cover a more extended period than the currently available Zubercal light curves. We utilize the median flux of the Zubercal light curves as the reference flux, a 3\arcsec-diameter aperture, and a 4\arcsec-8\arcsec\ radius annulus to subtract the local differential background. The difference image flux uncertainties are underestimated, so we use the mean intra-night variance as a systematic uncertainty, and add it in quadrature. We compare our difference image and Zubercal light curves in Fig.~\ref{fig:diff_image_lc}. The two light curves are consistent, with the difference image light curves producing a slightly lower mean flux uncertainty and intrinsic variability. In fact, the intrinsic variability for the three LRDs is even weaker using the difference image light curves than those based on the PSF light curves. 

We collected publicly available archival spectroscopy for the three local LRDs from SDSS, Keck, and Gemini. We reduced the Keck and Gemini spectra using \textsc{pypeit} \citep{pypeit:joss_pub}, following the cookbooks for both instruments. The spectral reduction steps include bias-subtraction, flat-fielding, wavelength calibration using arc lamps, sky subtraction, and flux calibration. We flux-calibrated the spectra using the companion observations of the standard stars Feige~66 (Gemini) and HZ~44 (Keck). We did not perform telluric corrections.

SDSS fiber-aperture spectra for the three local LRDs were taken in 2003--2005. The SDSS spectral resolution is $R \sim 1800$ at the \Ha\ line. Keck observations of J1025 and J1047 were taken in 2016 with the LRIS instrument (2016A C271LA; PI: F. Harrison). The red-side spectrum covers the \Ha\ line and was taken with the 400/8500 grating with a 1.5-arcsec slit width. The Keck spectral resolution with this setup is $R\sim 900$ at the \Ha\ line. High-resolution Gemini long-slit spectroscopy was obtained for J1025 and J1047 in 2020 using the GMOS instrument on Gemini-North (GN-2020A-FT-204; PI C. J. Burke). The setup was centered on the \Ha\ line with the R831 grating and a 1-arcsec slit width. The Gemini spectral resolution with this setup is $R\sim 2200$, but the effective spectral resolution is $R\sim 3400$ due to the size of the PSF compared to the slit \citep{Ji2025}.

\section{Results}\label{sec:results}

\begin{table*}[]
    \begin{tabular}{|c|c|c|c|c|}
        \hline
           \multicolumn{2}{|c|}{ \multirow{5}{7.5em}{} } & J1022+0841 & J1025+1402 & J1047+0739 \\
           \multicolumn{2}{|c|}{ \multirow{5}{7.5em}{} } & $z = 0.2227$ & $z = 0.1007$ & $z = 0.1682$ \\
        \hline
        \multirow{5}{7.5em}{$\sigma_0$ [mag]} & \emph{g} & ${ 0.118 }_{-0.041}^{+0.040}$ & ${ 0.043 }_{-0.012}^{+0.012}$ & ${ 0.018 }_{-0.018}^{+0.019}$ \\[2pt]
            & \textbf{\emph{r}} & $\mathbf{{0.000 }_{-0.000}^{+0.014}}$ & $\mathbf{{ 0.025 }_{-0.005}^{+0.004}}$ & $\mathbf{{ 0.020 }_{-0.005}^{+0.005}}$ \\[2pt]
            & \emph{i} & ${ 0.000 }_{-0.000}^{+0.015}$ & ${ 0.017 }_{-0.011}^{+0.008}$ & ${ 0.043 }_{-0.010}^{+0.011}$ \\[2pt]
            & W1 & ${ 0.057 }_{-0.016}^{+0.018}$ & ${ 0.034 }_{-0.010}^{+0.011}$ & ${ 0.023 }_{-0.012}^{+0.011}$ \\[2pt]
            & W2 & ${ 0.057 }_{-0.023}^{+0.023}$ & ${ 0.018 }_{-0.018}^{+0.020}$ & ${ 0.033 }_{-0.009}^{+0.010}$ \\[2pt]
        \hline
        \multirow{5}{7.5em}{$\overline{\sigma_i}$ [mag]} & \emph{g} & 0.204 & 0.103 & 0.108 \\
            & \textbf{\emph{r}} & 0.117 & 0.053 & 0.053 \\
            & \emph{i} & 0.070 & 0.037 & 0.052 \\
            & W1 & 0.073 & 0.053 & 0.040 \\
            & W2 & 0.099 & 0.074 & 0.046 \\
        \hline
        \multirow{2}{7.5em}{$(\Delta F / F)_{\rm total\ H\alpha}$} & Keck-SDSS & \nodata & $(-7.4^{+1.0}_{-0.4})\times10^{-2}$ & $(-7.3^{+2.6}_{-0.6})\times10^{-2}$ \\
            & {Gemini-SDSS} & \nodata & $(-2.7^{+1.0}_{-0.4})\times10^{-2}$ & $(3.3^{+2.6}_{-0.6})\times10^{-2}$ \\
        \hline
            \multirow{2}{7.5em}{$(\Delta F / F)_{\rm broad\ H\alpha}$} & Keck-SDSS & \nodata & $(-5.4^{+1.1}_{-0.8})\times10^{-2}$ & $(-4.6^{+2.4}_{-1.0})\times10^{-2}$ \\
            & \textbf{Gemini-SDSS} & \nodata & $\mathbf{(8.0^{+9.7}_{-7.9})\times10^{-3}}$ & $\mathbf{(5.9^{+23.3}_{-14.0})\times10^{-3}}$ \\
        \hline
    \end{tabular}
    
    \caption{Photometric and \halpha\ spectral variability measurements. The mean photometric uncertainty $\overline{\sigma_i}$ is shown for each filter. The total \halpha\ flux measurements may suffer from aperture effects for the narrow-line emission. }
    \label{tab:phot_properties}
\end{table*}

\subsection{Photometric variability}\label{sec:photo_var}

\begin{figure}
    \centering
    \includegraphics[width=\linewidth]{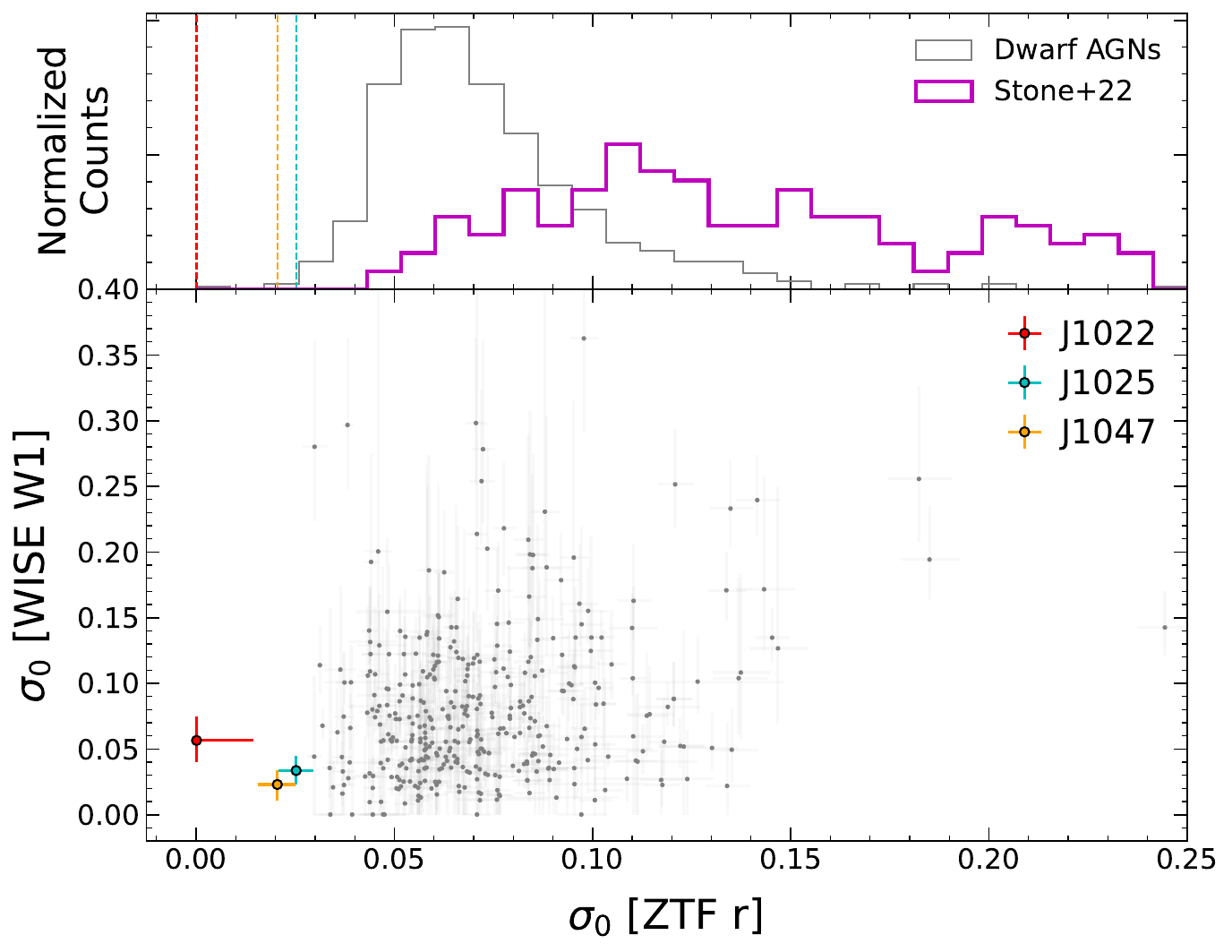}
    \caption{Intrinsic $r$-band variability measurements for different samples. \emph{Top:} The normalized $\sigma_0$ distribution in the \emph{r}-band for the dwarf AGN sample (gray) and a sample of broad-line quasars from \citet[][purple]{Stone_etal_2022}. The LRDs are shown as vertical dashed lines. \emph{Bottom:} The comparison of $\sigma_0$ in the ZTF \emph{r}-band and WISE W1 bands for the dwarf AGN sample (gray circles) and local LRDs (colored circles). }
    \label{fig:sigma0_comparison}
\end{figure}



In this work we focus on the $r$ band variability, which traces the rest-frame optical of these local LRDs. The three objects are unresolved in their $r$-band imaging \citep{Lin_etal_2025_egg}, indicating compact sizes of the rest-frame optical continuum emission. The point-source nature in $r$ band also makes the ZTF photometric measurements more reliable. For completeness we also report continuum variability measurements in other bands in Table~\ref{tab:phot_properties}, but caution that there may be underestimated systematics in these measurements. 

The light curves shown in Fig.~\ref{fig:lc} already suggest weak variability, as the magnitude fluctuations are roughly consistent with flux uncertainties. To measure the intrinsic variability from the ZTF light curves, we use a maximum-likelihood estimator \citep{Shen_etal_2019b} that treats each data point $X_i$ in the light curve as an independent measurement of a constant flux $\mu$ perturbed by the measurement uncertainty $\sigma_i$ and the intrinsic rms variability $\sigma_0$ (both processes are assumed to be Gaussian). The likelihood function can be written as:
\begin{equation}
    -2\ln L =\sum_{i=1}^N\frac{(X_i-\mu)^2}{\sigma_0^2+\sigma_i^2}+\sum_{i=1}^N\ln(\sigma_0^2+\sigma_i^2)\ .
\end{equation}

The ML estimator of the intrinsic rms variability $\sigma_0$ and its associated uncertainties are solved iteratively \citep[][and K. Horne, in prep]{Shen_etal_2019b}. The benefit of using the ML estimator is that it can deal with non-uniform flux errors across the light curve. While this simple ML estimator does not take into account potential dependence of intrinsic variability on timescale (e.g., a structure function), it provides a robust and efficient estimate of the rms variability over the time period of interest. 

Fig.~\ref{fig:sigma0_comparison} compares the derived intrinsic continuum variability $\sigma_0$ of the three local LRDs with the control dwarf AGN sample. The dwarf AGNs have significant host contamination in restframe optical, which reduces the measured fractional variability (rms magnitude). \citet{Burke_etal_2023} showed that after correcting for typical host galaxy contamination, the rms variability of dwarf AGNs is consistent with that for quasars. The local LRDs show significantly weaker continuum variability in rest-frame optical than normal dwarf AGNs and more massive SDSS quasars. The ML estimates of $\sigma_0$, while formally detected for J1025 and J1047, are at the $<3-4\%$ level (3$\sigma$) over a rest-frame time baseline of $\sim 5$~yrs. 

\begin{figure}
    \centering
    \includegraphics[width=.9\linewidth]{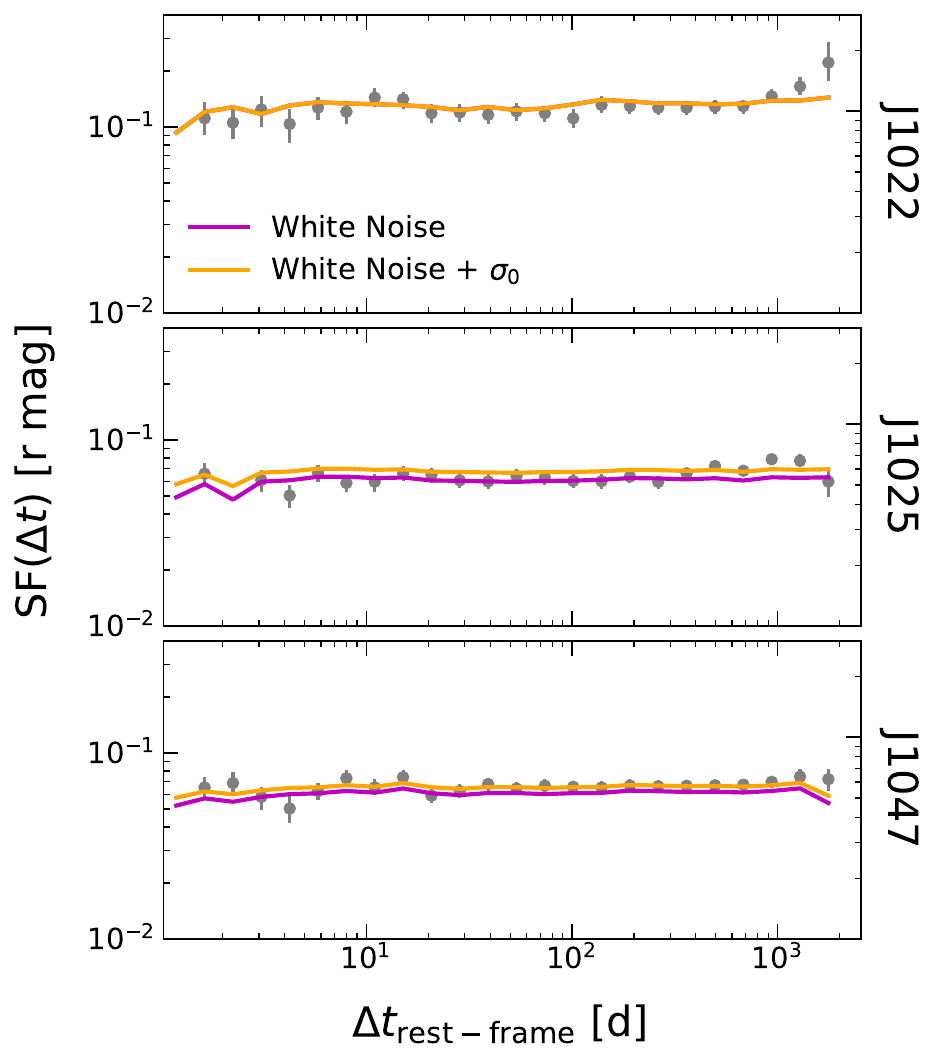}
    \caption{Structure function measurements for the three local LRDs in the \emph{r}-band. The structure functions are shown as gray circles, with uncertainties. The expected structure function assuming a pure white noise light curve is shown in purple. The expected structure function assuming a light curve generated by white noise and the measured intrinsic variability $\sigma_0$ is shown in orange. }
    \label{fig:structure_function}
\end{figure}

We further measure the structure function for the local LRDs and the comparison sample. We show results for the local LRDs in \emph{r}-band in Fig.~\ref{fig:structure_function}. The structure function measurement in each time-difference $\Delta t$ bin is the mean of magnitude differences $|\Delta m|$ in that bin, without correction for flux uncertainties. We measure the structure functions in the source rest-frame, and calculate uncertainties in the mean by bootstrapping with 500 realizations of each light curve. There is no discernible structure in the structure function of the local LRDs, although the overall weak intrinsic variability makes the SF measurements generally difficult and noisy. 

We test the reliability of these results by comparing the measured structure functions to expected structure functions generated by light curves from pure white noise. For each magnitude pair, we assume $\Delta m$ is distributed by a Gaussian with zero mean and dispersion determined by the uncertainty of the two magnitudes. We bin the expected values of $|\Delta m|$ in the same manner as the original structure function. We perform the same process, assuming the light curve is generated by white noise and a constant intrinsic variability $\sigma_0$. The measured structure functions for the three local LRDs are consistent with the latter prediction. On the other hand, the white-noise-like ``intrinsic'' variability could be interpreted as underestimated flux measurement uncertainties, which would indicate even lower intrinsic variability for these local LRDs.  

Finally, early $r$-band photometry for the three local LRDs from SDSS \citep{SDSS} and PanSTARRS-1 \citep{PS1} during 2002--2014 showed mild systematic offsets ($0.05-0.2$~mag, after accounting for filter transmission differences between SDSS $r$ and ZTF $r$) from the mean ZTF photometry. Within the single PanSTARRS-1 survey with 4--5 annual epochs, however, the rms $r$-band variability is less than a few percent, consistent with the ZTF constraints. The small mean offsets between different surveys are more difficult to interpret. It is possible that the mean flux varied by $\le 0.2$~mag over timescales much longer than rest-frame $\sim 5$~yrs as sampled by ZTF. On the other hand, the differences in photometric systems and zero-point uncertainties in ZTF make it challenging to confirm these systematic offsets across surveys for any individual objects. We therefore conclude that there is no evidence for optical continuum variability for these three local LRDs by more than a few percent between SDSS and ZTF.

\subsection{Spectral variability}\label{sec:spec_var}

\begin{figure*}
    \centering
    \includegraphics[width=0.48\linewidth]{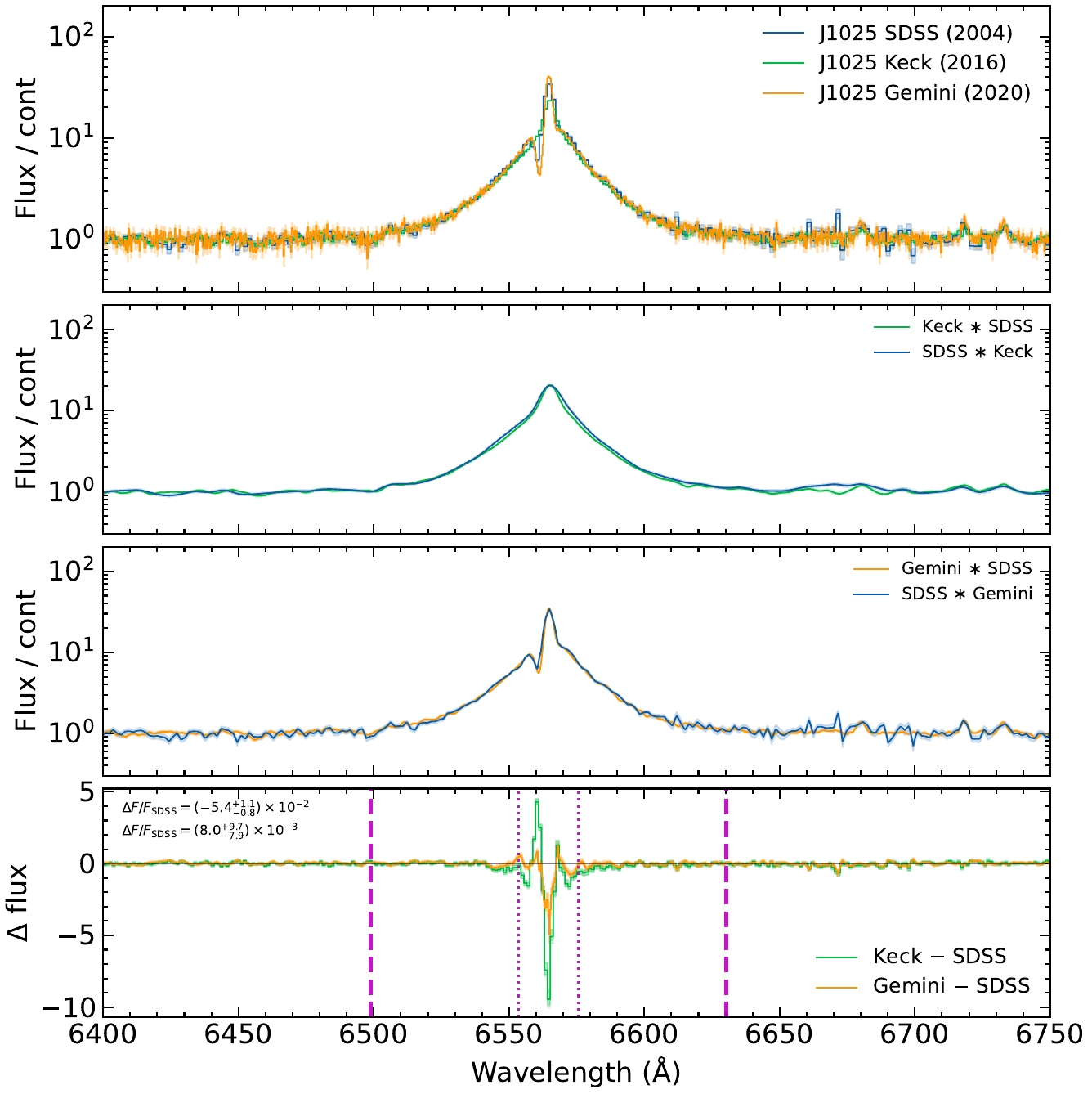}
    \includegraphics[width=0.48\linewidth]{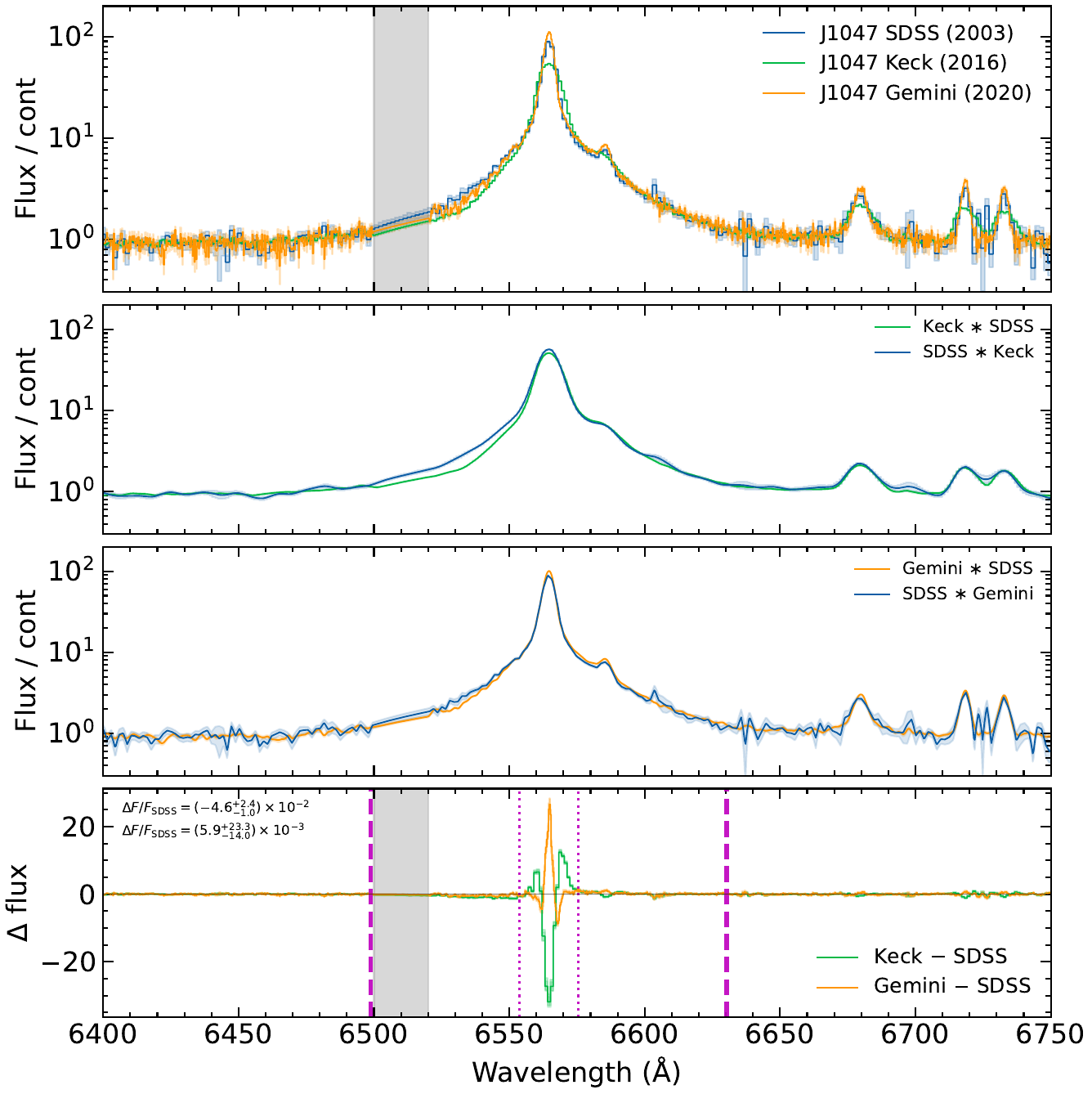}
    \caption{Multi-epoch \halpha\, spectroscopy over $\sim15$ years for J1025 (dubbed ``The Egg'' in \citet{Lin_etal_2025_egg}; left) and J1047 (right). The top panels show the reduced spectra for all three epochs. The middle panels show the matched-$R$ convolved spectra. The bottom panel shows the SDSS difference spectra. The gray bands in the right panel is the masked region due to telluric absorption.}
    \label{fig:Havar}
\end{figure*}

For spectral variability measurements we focus on J1025 and J1047 for which we have multi-epoch spectra. For each spectrum, we first transform it to the rest frame and fit a local linear continuum near the \Ha\ line. We performed a maximum likelihood fitting using the Levenberg-Marquardt method with an iterative clipping procedure, rejecting $3\sigma$ outliers. We used a 1-D polynomial model in the 6000--7000 \AA\ range, excluding a $\pm 100$ \AA\ window around the \Ha\ line. We normalized each spectrum by this local continuum to correct for slit losses, etc., as shown in the top panel of Figure~\ref{fig:Havar}. Because the continuum and broad \halpha\ emission are both point-like for the three LRDs, this normalization allows us to measure broad \halpha\ variability when the optical continuum is more or less constant (see Section~\ref{sec:photo_var}). 


To isolate instrumental differences, we then performed pair-wise resolution matching between instruments. In the middle panels of Figure~\ref{fig:Havar}, we convolved each spectrum to the instrumental resolution $R$ of the other instrument. We used a Gaussian kernel with width $\sigma_c = c / (2.35\, R)$. In the bottom panel, we then subtracted the convolved, interpolated, and continuum-subtracted Keck and Gemini spectra from the continuum-subtracted SDSS spectra. Finally, we integrated these residuals over a 3000 km s$^{-1}$ velocity window around \Ha. We tried this with and without excluding the narrow core ($\pm 500$ km s$^{-1}$) to measure the fractional flux difference $\Delta F/F_{\rm SDSS}$. To estimate the uncertainties, we perform 100 Monte Carlo random samples from the flux error spectra and calculated the 16th and 84th percentile ranges as approximate $1\sigma$ uncertainty estimates. 

The measured fractional change is at the few percent level for the total \halpha\ flux, and is lower for the broad \halpha\ flux. Since the narrow-line flux could be extended beyond the spectral aperture, we consider the constraints on the broad-line flux fractional changes more reliable. The Keck-SDSS difference is larger than the Gemini-SDSS difference for both J1025 and J1047. However, there is no reason to expect that both objects would vary in sync, and return approximately to the SDSS state during the Gemini epoch. Therefore there are unknown systematics in the Keck spectral reduction, possibly due to the significantly lower spectral resolution of the Keck spectrum compared with SDSS and Gemini. The Gemini-SDSS differential spectra suggest fractional variability of less than a few percent for both J1025 and J1047, over a rest-frame time span of $\sim 15$~yrs. If the underlying optical continuum had varied by more than a few percent over this period, it would be difficult to understand why the continuum-normalized broad \halpha\ flux (technically the equivalent width) would remain the same, since the broad \halpha\ emission is not powered by the optical continuum. 

Finally, we measured non-parametric emission-line widths and equivalent widths (EWs) and full width at half-maxima (FWHMs) directly from the continuum-subtracted spectra after masking the narrow line, without assuming any specific line profile model. The FWHMs were determined by locating the half-maximum crossings on either side of the line peak in the flux profile. For line strengths, we computed the equivalent width by numerically integrating the flux excess relative to the local continuum level across the line, following
\begin{equation}
\mathrm{EW} = \int \frac{F_{\lambda} - F_{\mathrm{cont}}}{F_{\mathrm{cont}}} \, d\lambda.
\end{equation}

This non-parametric approach provides robust, model-independent measurements of line width and strength.

\subsection{Broad \halpha\ properties}

\begin{figure}
    \centering
    \includegraphics[width=\linewidth]{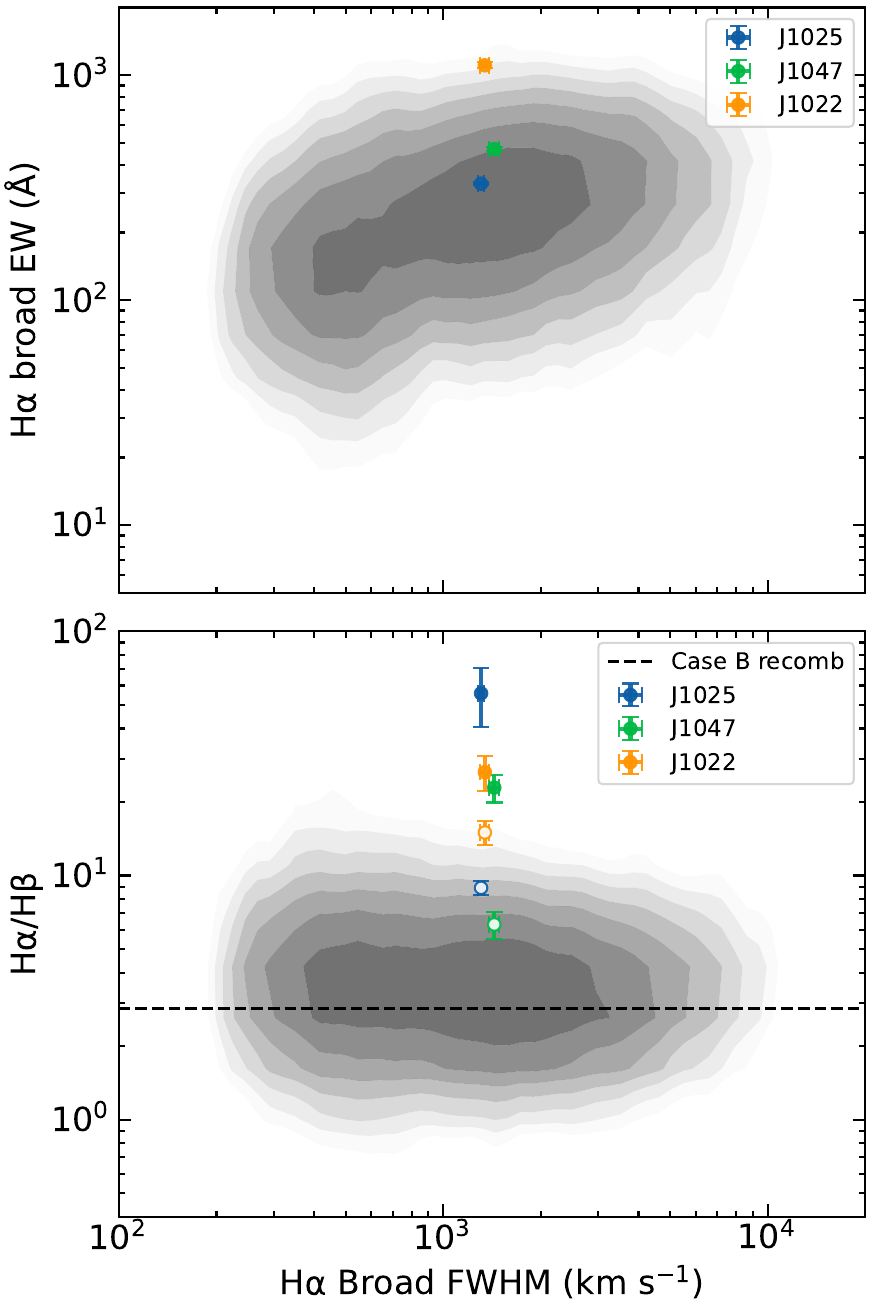}
    \caption{Comparison between broad Balmer line properties for the local LRDs (colored points) and broad-line quasars (gray contours) from SDSS \citep{Wu_DR16Q}. \textbf{Top:} LRDs have systematically higher broad \halpha\ EW and deviate significantly from the average population of broad-line AGNs. \textbf{Bottom:} The \halpha/\hbeta\ ratio for local LRDs far exceeds the theoretical value for Case B recombination (the dashed horizontal line). Because LRDs have little dust extinction, the large \halpha/\hbeta\ ratio suggests the lines are primarily collisionally excited. The solid symbols correspond to the broad line flux ratio and the open symbols correspond to the total line flux ratio.}
    \label{fig:fwhm_ew}
\end{figure}

We show the distribution of the three local LRDs in the broad \halpha\ FWHM versus restframe equivalent width (EW) plane in Fig.~\ref{fig:fwhm_ew}. The local LRDs have systematically larger EW compared with the average broad-line AGN population. While this deviation by itself does not preclude the use of single-epoch virial mass estimators based on broad \halpha\ to estimate BH masses in LRDs, it does provide a cautionary note on the applicability of these mass scaling relations to the LRD population. 

Fig.~\ref{fig:fwhm_ew} also compares the \halpha/\hbeta\ flux ratios between the LRDs and normal broad-line AGNs from \citet{Wu_DR16Q}. We measure total and broad \halpha/\hbeta\ ratios as $15.0\pm1.7$ and $26.5\pm4.3$ (J1022), $8.9\pm0.6$ and $55.7\pm15.0$ (J1025), $6.3\pm0.8$ and $22.9\pm3.0$ (J1047). In decomposing the broad and narrow components, we fix the velocity widths and offsets of the narrow-line components to those of the \SII\ line. 

The three local LRDs have exceptionally larger \halpha/\hbeta\ ratios, far exceeding the Case B value. While dust extinction can potentially explain the large \halpha/\hbeta\ ratios, the three local LRDs are known to have little dust emission, ruling out a large Balmer decrement due to dust. These large \halpha/\hbeta\ ratios are then indicative of collisionally excited lines \citep[e.g.,][]{Ji2025,Torralba_2025,Nikopoulos_2025} in the three local LRDs. We discuss this possibility further in Section~\ref{sec:disc}.


\section{Discussion}\label{sec:disc}

We now discuss the variability properties of local LRDs in the context of recent models for high-redshift LRDs. In particular, the semi-analytic model in \citet{LiuEtAl2025} postulates that the rest-frame optical continuum of LRDs originates from the black body emission of a photosphere surrounding the accreting SMBH, with an effective temperature of $\sim 5000\,$K. The radius of the photosphere is roughly $10^4-10^5\,r_{g}$ or $\sim 10^3\,{\rm AU}$ for a $10^6\,M_\odot$ black hole. The Keplerian speed at this radius is roughly $\sim 2000\,{\rm km\,s^{-1}}$, comparable to the observed broad-line width. In this model, the luminosity of the photosphere is likely capped at $\sim L_{\rm Edd}$, while the mass accretion rate on to the black hole can significantly exceed $\dot{M}_{\rm Edd}$ \citep[e.g.,][]{Jiang_etal_2025}. 

\subsection{Optical continuum variability}

The continuum variability of normal AGNs is known to decrease with Eddington ratio \citep[e.g.,][]{Ai_etal_2010,Macleod_etal_2010,Rumbaugh_etal_2018}. For $L/L_{\rm Edd}\gtrsim 1$ AGNs (e.g., narrow-line Seyfert 1s, NLS1s), the average rms variability can reach as low as $\sim 5\%$ \citep{Ai_etal_2010}. Our inferred rest-frame optical continuum variability of $<3-4\%$ would be consistent with the extrapolation of the variability-Eddington ratio relation found for normal AGNs to even higher Eddington ratios. However, for normal broad-line AGNs, the Eddington ratio on average increases with increasing broad optical \FeII\ strength and decreasing \OIII\ strength, known as the Eigenvector 1 relations \citep[][]{Boroson_Green_1992,Shen_Ho_2014}. The local LRDs do not seem to follow the Eigenvector 1 relations with extreme properties of strong \FeII\ and weak \OIII\ emission compared with \hbeta\ \citep{Lin_etal_2025_egg}. While there are several similarities between LRDs and the overall NLS1 population \citep[e.g., see the recent review by][]{Berton_etal_2025}, many peculiar properties (such as the Balmer absorption and the rest-frame optical SED) are unique to LRDs and require different accretion models.  


Rigorous predictions of optical continuum variability from the photosphere emission in super-Eddington accreting SMBHs are not yet available. For example, many theoretical studies on high-Eddington accretion flows suggest low optical variability \citep[e.g.,][]{Inayoshi&Maiolino2025}, but no quantitative predictions are given. \citet{SecundaEtAl2025} used an empirical prescription to predict the continuum variability from super-Eddington SMBHs. In their prescription, they adopted the predicted UV light curves in the inner accretion region (deep within the photosphere) from radiation MHD simulations in \citet{Jiang_etal_2025} for super-Eddington AGNs. They then extrapolate the variability structure function fits to the UV light curve to the optical emitting region at $r\sim 10^4-10^5\,r_{g}$ to predict the optical variability. This empirical prescription is obviously simplified, and ignores the detailed radiative transport to the photosphere. Nevertheless, \citet{SecundaEtAl2025} predict an optical rms variability of $\sim$few percent on rest-frame yearly timescales. Our measured optical variability for local LRDs is slightly lower than this prediction, but generally consistent. 

Photometric variations of the photosphere emission from super-Eddington accreting SMBHs are likely to occur on the dynamical timescale at the photosphere \citep[e.g.,][]{LiuEtAl2025}. 
The photosphere radius is around $r_{\rm ph}\approx 10^4-10^5\,r_{\rm g}$. At $r_{\rm ph}$, the Keplerian circular speed $V_{\rm ph}$ and dynamical timescale $t_{\rm dyn,ph}$ are: $V_{\rm ph}=\sqrt{GM/r_{\rm ph}}=(r_{\rm ph}/r_g)^{-1/2}c\approx 1000-3000\,{\rm km\,s^{-1}}$, and $t_{\rm dyn,ph}\approx r_{\rm ph}/V_{\rm ph}=(r_{\rm ph}/r_g)^{3/2}(r_g/c)\approx 1.6\times 10^{-7}{\rm yr}(M_{\rm BH}/10^6\,M_\odot)(r_{\rm ph}/r_g)^{3/2}$. For a fiducial $M_{\rm BH}=10^6\,M_\odot$ black hole, the expected rest-frame dynamical timescale at the photosphere is therefore $\sim 0.16-5$\,yr if assuming $r_{\rm ph}/r_{g}\approx 10^4-10^5$ \citep{LiuEtAl2025}. Alternatively, one can assume that the black body emission from the photosphere is capped at some Eddington ratio $\lambda_{\rm Edd}=4\pi r_{\rm ph}^2\sigma_{\rm SB}T_{\rm eff}^4/L_{\rm Edd}$ and derive the required photosphere radius \citep[e.g.,][]{Inayoshi_etal_2025b}. For $\lambda_{\rm Edd}=1$, $M_{\rm BH}=10^6\,M_\odot$ and $T_{\rm eff}=5000$\,K, the expected dynamical time $t_{\rm dyn,ph}\approx r_{\rm ph}/V_{\rm ph}\approx 6$~yrs. {If the photosphere is in the outflow region with terminal speed $v_w$ as commonly found for super-Eddington accretion systems \citep{Jiang+2019,Jiang_etal_2025}, another relevant timescale for variability is the photon diffusion time from the wind $t_{\rm diff,w}\approx r_{\rm ph}\tau_w/c$, where the relevant optical depth $\tau_w=c/v_w$, which means $t_{\rm diff, w}\approx r_{\rm ph}/v_w$. The terminal speed $v_w$ of radiation driven wind can be smaller than the local Keplerian speed by a factor of a few, which means the photon diffusion time will be longer than the local dynamical time by another factor of few.} These order of magnitude estimates suggest that the observational baseline for the three local LRDs explores similar timescales comparable to the dynamical timescale of the hypothesized photosphere. 

The observed low optical continuum variability in the three local LRDs disfavors scenarios where the photosphere of the super-Eddington accreting SMBH experiences large-amplitude (e.g., $\gg$ a few percent), opacity-driven pulsations \citep[e.g.,][]{Inayoshi_etal_2025b}, reminiscent of the periodic variations of Cepheids. However, low-amplitude modes in such LRD photospheres can still be produced, which would require more precise measurements of the continuum light curve. If we simply fit a sinusoidal function to the difference image light curves for the three LRDs in Fig.~\ref{fig:diff_image_lc}, the best-fit amplitude is at the few percent level at most, and is formally non-detection. Long-term monitoring of these LRDs sampling different phases is necessary to robustly constrain variability patterns, e.g., to measure the variability power spectrum.  

It has also been proposed that LRDs are ``quasi-stars" with a black hole at the center and a massive envelope in the outer part \citep{Begelman_etal_2008,Begelman_Dexter_2025} {reaching $10^5-10^6\ r_g$.} The outer envelope is typically constructed following the same way as the static structure of the envelope of massive stars, even though the mass of these quasi-stars is many orders of magnitude larger. Therefore, scaled stellar spectra with effective temperatures around $5000$~K can be directly applied to explain the observed spectra of LRDs. Recent 1D models of quasi-stars based on MESA \citep{Hassan+2025,Santarelli+2025} typically find that saturated convection is needed to transport the luminosity out to avoid the envelope to be blown away and density inversion can show up near the photosphere due to strong radiation force caused by opacity peaks, which is a well-known properties of massive star envelopes in 1D \citep{Paxton+2013}. However, these 1D structures with density inversion in massive star envelopes are unstable in 3D and typically result in strong variability of the light curves as found by 3D radiation hydrodynamic simulations \citep{Jiang+2015,Jiang+2018,Jiang2023}, which can be used to explain the commonly observed low frequency variability of massive stars by TESS \citep{Schultz+2022}. If the quasi-star models also show similar unstable envelope structures in 3D, they can produce significant variability on the local dynamical timescales at the location of the density inversion, which may not be consistent with variability observations. Consequently, long-term variability measurements can be used to place strong constraints on quasi-star models. 

\subsection{Broad \halpha\ variability}

The interpretation of broad \halpha\ line variability is more complicated. The measured fractional broad \halpha\ variability is at the $\sim$few percent level at most (Section~\ref{sec:spec_var}). Under the framework of the super-Eddington accretion photosphere model \citep[]{LiuEtAl2025}, \citet{Lin_etal_2025_egg} postulated that the broad line clouds are photoionized by leaked EUV radiation in the polar region. In this scenario, the variability in broad \halpha\ should trace the variability in the UV emission from the inner accretion flow. The BLR region is rather compact (i.e., near the photosphere) and the light-crossing time is only a few days, leading to negligible geometric dilution of the driving UV variability. 
The UV light curves in the simulations of super-Eddington accretion in \citet{Jiang_etal_2025} display rms variability of $\sim 10\% - 20\%$ on rest-frame yearly timescales, and roughly follow a power-law power density distribution with reduced variability at shorter timescales. If broad \halpha\ is photoionized by the leaked UV flux, we expect to see similar variability amplitude in the broad \halpha\ over a few years in restframe.


The broad \halpha\ profiles in LRDs are often observed to display exponential wings, characteristics of electron scattering in line broadening from photoionized gas \citep[e.g.,][]{Weymann_1970,Rusakov_etal_2025,Chang_etal_2025}. In the dense gas enshrouding models, neutral hydrogen scattering could also be an important contributor to extra line broadening \citep[e.g.,][]{Naidu2025,Chang_etal_2025}. Because the hydrogen density is very high \citep[$n_{\rm H}\sim 10^{9-10}\,{\rm cm^{-3}}$, e.g.,][]{Torralba_2025}, and if most of the UV ionizing flux is attenuated by the optically thick inner gas, the Balmer-emitting region will not be as highly ionized as the broad-line region gas in normal AGNs. Under such circumstances, collisional excitation may become the primary mechanism to produce the broad \halpha\ emission. This scenario is consistent with the observed large \halpha/\hbeta\ ratios in the three LRDs, as well as the lack of significant line variability over decades.

\citet{Torralba_2025} performed a more quantitative calculation for the line emission in a $z=5.5$ LRD, under the general framework of the BH* model \citep{Naidu2025}. They also found a large \halpha/\hbeta\ ratio of $\sim 10$ and concluded that the Balmer lines are primarily produced by collisional excitation and resonant scattering within a warm layer inside a BH*. They also suggested that the broad wings of the line are due to electron scattering \citep[e.g.,][]{Rusakov_etal_2025} and thus do not reflect the virial velocity of the line-emitting region. 

While there are still qualitative and quantitative differences among these recent LRD models (quasi-star, BH*, super-Eddington accretion photosphere), their commonalities of high gas density, low dust extinction, and thermalized atmosphere emission can explain most of the observed LRD properties. The BH* model \citep[e.g.,][]{Naidu2025,Torralba_2025} is currently the only quantitative model to predict emission line properties. Under the BH* model, the broad \halpha\ line cannot be used as a reliable indicator for viral black hole masses, and the theoretical masses required in these models are typically orders of magnitude lower than that based on broad \halpha. Our variability and spectral measurements of the three local LRDs are in support of this scenario.      





\subsection{Other scenarios}

Alternative models for LRDs are available \citep[e.g.,][]{Baggen2024,Bellovary2025,Zhang2025sed} that do not feature the dense gas-enshrouded SMBH scenario. However, these models are difficult to explain the myriad of LRD properties. For example, the tidal disruption model \citep{Bellovary2025} is inconsistent with the lack of significant variability in most LRDs. The extremely dense stellar population model \citep{Baggen2024} also has difficulties to explain the overall rest-frame optical-IR SED and the exceptionally strong Balmer breaks observed in LRDs. Alternative SMBH accretion models invoking composite accretion disks \citep[e.g.,][]{Zhang2025sed} could simultaneously explain the rest-optical emission and the rest-UV upturn. However, recent imaging analyses of the rest-UV emission of LRDs show that the UV-emitting region is spatially resolved ($\gtrsim$ tens of parsec) and not originated from nuclear emission \citep[e.g.,][]{Chen+2025a,ZhuangEtAl2025a}, disfavoring the composite accretion disk scenario. 


\section{Conclusions}\label{sec:con}

The recently proposed connection between some local metal-poor dwarf AGNs and high-redshift LRDs by \citet{Lin_etal_2025_egg} enabled a unique opportunity to study the long-term variability of these peculiar systems. In this work we use public light curves and multi-epoch spectra of the three local LRDs (J1022, J1025 and J1047) in \citet{Lin_etal_2025_egg} to constrain their rest-frame optical continuum and broad \halpha\ variability. 

With ZTF public light curves that span a rest-frame timescale of $\sim 5$~yrs, we were able to derive stringent constraints on the intrinsic optical continuum variability of the three local LRDs, which vary by less than $3-4\%$ over this time period. Additionally, using multi-epoch spectra of J1025 and J1047 that span $\sim 15$~years in their rest-frame, we place constraints on the fractional broad \halpha\ variability to be less than a few percent. These are by far the strongest constraints on the variability properties of local LRD analogs on multi-year to decade timescales, and are consistent with the majority of variability constraints on much shorter timescales \citep[$\lesssim 1$~yr in restframe, e.g.,][]{Kokubo2024,ZhangEtAl2025,Stone_etal_2025}.

Our variability results, as well as the exceptional spectral properties of broad \halpha\ compared with normal broad-line AGNs, are generally consistent with recent LRD models that feature a dense, cool ($T_{\rm eff}\approx 5000\,$K) photosphere enshrouding an accreting SMBH. The weak (or lack of) long-term continuum variability provides strong constraints on the detailed structure and stability of such models. 

On the other hand, the lack of significant broad \halpha\ variability, large \halpha/\hbeta\ ratios and little dust extinction \citep[e.g.,][]{Setton_etal_2025,Xiao_etal_2025}, are inconsistent with the line being powered by photoionization. Instead, we suggest that the broad Balmer lines are  collisonally excited, which is consistent with the high gas density of recent LRD models and more quantitative calculations for high-redshift LRDs \citep[e.g.,][]{Torralba_2025}. In such a scenario, there is no need for the broad Balmer lines to be around the polar region of the photosphere to receive leaked ionizing flux \citep{Lin_etal_2025_egg}. The latter picture may result in an orientation dependence of observed LRD properties, which can be further tested with more LRD observations. A general conclusion, as already suggested in earlier work \citep[e.g.,][]{Naidu2025,Torralba_2025}, is that broad \halpha\ can no longer be used as a reliable indicator for the black hole mass. This is because not only the scaling relations for virial black hole masses are based on the assumption that the line is photoionized, but also electron scattering and hydrogen resonant scattering in such dense environment can significantly alter the broad-line profiles. 

Continued monitoring of these local LRD analogs with better photometric and spectroscopic precision will further tighten these variability measurements, and perhaps reveal variability on even longer timescales. Given the brightness of these objects and the point-source nature, we should be able to reach $\lesssim 10$~mmag photometric precision -- for comparison, ZTF photometric precision for these objects is at the 0.05--0.1~mag level. This would measure the intrinsic continuum variability to even higher precision. On the other hand, continued spectroscopic monitoring can further confirm the broad line variations over time, which would place stronger constraints on models. 

\begin{acknowledgements}

We thank Keith Horne for help with the implementation of the maximum-likelihood estimator, Zijian Zhang, Fengwu Sun, and Rohan Naidu for useful discussions. CJB is supported by an NSF Astronomy and Astrophysics Postdoctoral Fellowship under award AST-2303803. This material is based upon work supported by the National Science Foundation under Award No. 2303803. This research award is partially funded by a generous gift of Charles Simonyi to the NSF Division of Astronomical Sciences. The award is made in recognition of significant contributions to Rubin Observatory’s Legacy Survey of Space and Time. The ztfquery code was funded by the European Research Council (ERC) under the European Union's Horizon 2020 research and innovation programme (grant agreement n°759194 - USNAC, PI: Rigault). The Center for Computational Astrophysics at the Flatiron Institute is supported by the Simons Foundation.
\end{acknowledgements}

\software{
\texttt{Astropy} \citep{2013A&A...558A..33A,2018AJ....156..123A,astropy_3}, 
\texttt{Matplotlib} \citep{Hunter2007}, 
\texttt{Numpy} \citep{Harris2020}, 
\texttt{photutils} \citep{photutils}, 
\texttt{scipy} \citep{scipy},
\texttt{ztfquery} \citep{ztfquery}
}

\bibliography{sample631, ZRefs}{}
\bibliographystyle{aasjournal}


\end{CJK*}
\end{document}